\documentclass{PoS}
\usepackage{amsmath}
\usepackage{appendix}
\usepackage{mathtools}
\usepackage{tikz}
\usepackage{cite}

\newcommand{\DefOfF}[0]{Luscher:1990ux,Rummukainen:1995vs,Kim:2005gf}
\newcommand{\ThreeTransitions}[0]{Muller:2020wjo,Hansen:2021ofl}

\title{Higher partial wave contamination in finite-volume formulae for 1-to-2 transitions}

\ShortTitle{}

\author{Maxwell T. Hansen, \speaker{Toby Peterken}\\
Higgs Centre for Theoretical Physics, School of Physics and Astronomy, The University of Edinburgh, Edinburgh EH9 3FD, UK\\
E-mail:
\email{maxwell.hansen@ed.ac.uk},
\email{t.peterken@sms.ed.ac.uk}}

\abstract{It is common practice to truncate the finite-volume formula for $K\to\pi\pi$, and other one-to-two transitions, to only include the lowest partial wave, as in the original derivation by Lellouch and L{\"u}scher. However, as the precision of lattice calculations increases, it may become important to assess the systematic uncertainty of this approximation. With this motivation, we compare the $S$-wave-only ($\ell=0$) results with those truncated at the next lowest value of angular momentum that can contribute.}

\FullConference{The 38th International Symposium on Lattice Field Theory, LATTICE2021, 26th-30th July, 2021\\Zoom/Gather@Massachusetts Institute of Technology}

\begin{document}

\section{Introduction}
In the early 2000s, Lellouch and L{\"u}scher derived a formalism to extract $K \to \pi \pi$ decay amplitudes from finite-volume matrix elements calculable in lattice QCD \cite{Lellouch:2000pv}. This method has since been applied by the RBC-UKQCD collaboration (see e.g.~refs.~\cite{Bai:2015nea,Blum:2015ywa,Abbott:2020hxn}) in a first-principles calculation of the $K \to \pi \pi$ amplitudes with both allowed final states (isospin $0$ and $2$), leading to a first-principles understanding of the $\Delta I = 1/2$ rule and a determination of the CP violating parameter $\epsilon'/\epsilon$.

The original derivation of Lellouch and L{\"u}scher assumes vanishing spatial momentum in the finite-volume frame, and also that the effect of the $\ell=4$ (and higher) partial waves in $\pi\pi\to\pi\pi$ scattering can be neglected. Even though the incoming kaon can only couple to an $S$-wave two-pion final state, neglecting the $\ell = 4$ component amounts to an approximation in the value of the conversion factor relating the finite-volume matrix element to the infinite-volume decay amplitude. In addition, the original derivation only applies when the $\pi \pi$ finite-volume state, used to construct the matrix element, is sufficiently far below from the 9th state (8th excited state) of the system. Because the final-state energy must match the incoming kaon mass, $m_K$, this translates to a constraint that the volume be sufficiently small for the 8th excited state to sit above $m_K$. As described in ref.~\cite{Lellouch:2000pv}, the subtlety arises because the 8th excited state exhibits an accidental degeneracy that invalidates the original derivation, and as we show in this work, this is in fact closely tied to the role of angular-momentum truncation.

Subsequently, a series of publications has lifted these assumptions to provide a generic framework for extracting $0 \overset{\mathcal J}{\to} 2$ and $1 \overset{\mathcal J}{\to} 2$ transitions from finite-volume information~\cite{Lin:2001ek,Detmold:2004qn,Kim:2005gf,Christ:2005gi,Meyer:2011um,Hansen:2012tf,Briceno:2012yi,Bernard:2012bi,Agadjanov:2014kha,Briceno:2014uqa,Feng:2014gba,Briceno:2015csa}. The formalism at present holds for any number of two-particle channels,\footnote{Generalizations for transitions with three-particle final states have also been recently derived \cite{\ThreeTransitions}.} including any desired angular-momentum truncation. The purpose of this work is to consider the effect of the $\ell=4$ partial wave in the original context of $K \to \pi \pi$ decays. As computing power increases and lattice calculations become more precise, the contamination from $\ell=4$ could become relevant. Similarly, with lattice volumes as they are at present, the kaon mass is safely below the 8th energy level. But again, as computing resources increase and algorithms improve, there is potential for studies with lattice volumes large enough that accidentally degenerate states play a role.

This proceedings is organised as follows: Section~\ref{sec:luscher} recalls the original Lellouch-L{\"u}scher formalism. Section~\ref{sec:hansen} presents the generalization of ref.~\cite{Briceno:2014uqa} in the context of $K \to \pi \pi$, and section~\ref{sec:l0} reviews an exercise, presented in the same reference, in which one truncates the general formalism at $\ell=0$ and recovers the Lellouch-L{\"u}scher result. Then, in section~\ref{sec:l4}, we explicitly calculate the effect of the next lowest angular momentum state, $\ell=4$ for two pions with zero total momentum in a cubic, periodic finite volume. Finally, section~\ref{sec:accdeg} considers the general formalism in the vicinity of the 9th energy level in which we find a qualitative difference between the $\ell=0$ and $\ell=4$ truncations. Details not included here, such as a detailed analysis of the results for physical scattering parameters and the role of different boundary conditions and non-zero spatial momentum, will be presented in the full manuscript, to appear.

\section{Recap of the Lellouch-L{\"u}scher formalism}
\label{sec:luscher}

We denote the desired infinite-volume amplitude by $A(E)$:
\begin{equation}
A(E) \equiv \langle \pi \pi, E \vert \mathcal L_W(0) \vert K \rangle \,,
\end{equation}
where $\mathcal L_W(0)$ is the Lagrangian density mediating the transition, $\langle \pi \pi, E \vert$ is an infinite-volume, $S$-wave, two-pion out-state with energy $E$ and vanishing spatial momentum, and $\vert K \rangle$ is a kaon state. Both incoming and outgoing states have the standard relativistic normalization. Here we include a slight extension relative to ref.~\cite{Lellouch:2000pv} by allowing the final state energy to differ from the incoming kaon. Strictly this extension is also due to later work, in particular refs.~\cite{Meyer:2011um,Briceno:2014uqa}.

The finite-volume matrix element that can be extracted on the lattice is
\begin{align}
M_{n,L} =\langle \pi\pi, n, L| H_{W}|K, L\rangle, \qquad H_W=\int d^3x\ \mathcal{L}_W(x)=L^3\mathcal{L}_W(0) \,,
\end{align}
where both incoming and outgoing states are normalised to unity and the last equality has used translation invariance to compute the integral. (This uses the fact that the operator is inserted between states with matching spatial momenta.) The relation connecting $A(E)$ and $M_{n,L}$ is \cite{Lellouch:2000pv}
\begin{align}
\label{eq:LL_form}
\big |A(E_n) \big |^2 = 8\pi \left\{q \frac{\partial \phi}{\partial q}+p\frac{\partial \delta_0}{\partial p}\right\}_{p=p_n} \frac{m_K E^2}{p_n^3} |M_{n,L}|^2\bigg\vert _{E=E_n} \,,
\end{align}
where $q$ is the dimensionless version of the momentum $p$, defined as $q = p L/(2 \pi)$, and $p_n$ is the back-to-back pion momentum evaluated at a particular finite-volume energy, where the kinematic quantities are defined as
\begin{align}
p^2_n = E_n^2/4-m^2_\pi \,.
\end{align}
Here $\delta_0$ is the $S$-wave scattering phase shift for $\pi\pi\to\pi\pi$ scattering and $\phi$ is a known function that encodes the finite-volume effects and includes no dynamical information. (This function can be found in eq.~(A.1) of appendix B of ref.~\cite{Lellouch:2000pv} and is also defined in this work, implicitly, by combining eqs.~\eqref{eq:phidef} and \eqref{eq:fdef}.)

In order to interpret this physically as a $K\to\pi\pi$ decay, the finite-volume lattice energy used must coincide with the kaon mass, $E_n = m_K$. In fact this assumption was built in to the derivation of ref.~\cite{Lellouch:2000pv} but subsequently relaxed so that one can extract $A(E)$, which is perfectly well-defined away from physical kinematics, from the lattice calculation. The original formalism makes a few additional assumptions:
\begin{itemize}
\item The total momentum of both the incoming kaon and the outgoing two-pion state is zero in the finite-volume frame,
\item The scattering phase shift for $\pi\pi\to\pi\pi$ is negligible for even partial waves $\ell\geq 4$\,,
\item $E_n$ is one of the 8 lowest finite-volume energy levels.
\end{itemize}

The rest of this proceedings looks at the result of relaxing the final two assumptions, while keeping the first in place.\footnote{The full paper, to appear, will also address the first assumption by considering the effect of non-zero spatial momenta in the finite-volume frame.}

\section{The generalized $1\to2$ formalism}
\label{sec:hansen}

Reference~\cite{Briceno:2014uqa} provides a generalised $1 \to 2$ formalism by relaxing all of the restrictions listed above and more. This allows one to extract $1 \to 2$ transition amplitudes for systems with multiple coupled two-particle channels, any value of spatial momentum and, most importantly for this work, with any desired truncation in the angular momenta that contribute.

The formalism of ref.~\cite{Briceno:2014uqa} is most conveniently discussed in terms of the K-matrix. Taking $\delta_{\ell}(p)$, the $\pi \pi \to \pi \pi$ scattering phase in the $\ell$th partial wave, as a known entity, we define the K-matrix $\mathcal K$ and a closely related object $\widetilde{\mathcal K}$ as follows:
\begin{align}
\widetilde {\mathcal K}_{\ell' m', \ell m}(E) \equiv \delta_{\ell' \ell} \delta_{m' m} \frac{1}{p^{2 \ell}} \mathcal K^{(\ell)}(E)
\,, \qquad \text{where} \qquad \mathcal K^{(\ell)}(E) = \frac{16 \pi E}{p} \tan \delta_{\ell}(p) \,.
\end{align}
We then introduce the matrix
\begin{align}
\mathbb M(E,L)\equiv\widetilde {\mathcal{K}}(E)+\widetilde {F}(E,L)^{-1} \,,
\end{align}
where $\widetilde {F}$ is a known function that depends only on physical kinematics and the geometry of the finite volume. An explicit form is given in the appendix.

The matrix $\mathbb M$ serves two purposes. First, the condition that its determinant should vanish defines a quantisation condition equivalent to that of refs.~\cite{Luscher:1990ux,Rummukainen:1995vs,Kim:2005gf}. That is, at fixed $L$ and up to corrections falling as $e^{- m_\pi L}$, the values of $E$ satisfying
\begin{align}
\text{Det}\left[\mathbb M(E,L)\right]=\text{Det}\left[\widetilde {\mathcal{K}}(E)+\widetilde {F}(E,L)^{-1}\right]=0 \,,
\end{align}
are the finite-volume energies $E_n$ of the system. Second, the matrix allows one to give the updated formalism of ref.~\cite{Briceno:2014uqa}. Focusing here on a single flavour channel of two identical scalars (i.p.~two pions), the updated formalism for $1\to2$ transitions is
\begin{align}
\big |A(E_n) \big |^2 =2 m_K \mathcal{C}(E_n,L)\ |M_{n,L}|^2 \,,
\end{align}
where
\begin{align}
\label{eq:conv_fact}
\mathcal{C}(E,L)\equiv \frac{ \cos^2 \delta_0 (E) }{\text{Det}\left[\mathbb M\right]\left[\mathbb M^{-1}\right]_{00}}\,
\frac{ \partial\ \text{Det} \big [ \mathbb M(E, L) \big ] }{\partial E}.
\end{align}
The only difference between these expressions and the Lellouch-L{\"u}scher result is the definition of the conversion factor $ \mathcal{C}(E,L)$.

\section{Contribution from $\ell=0$ scattering}
\label{sec:l0}
In principle, $\mathbb M,\ \widetilde{\mathcal{K}}$ and $\widetilde{F}$ are infinite-dimensional matrices and so they need to be truncated in some way. In this section we make the assumption that the $\ell \geq 4$, $\pi\pi\to\pi\pi$ partial waves are all negligible (formally we set these to be identically zero). The quality of this approximation is therefore dictated by the size of the $\ell=4$ partial wave. Truncating the matrices at $\ell=0$ we have
\begin{align}
\mathbb M(E, L) &\equiv
\mathcal K^{(0)}(E) + \widetilde F_{00}(E, L)^{-1} ,\\
\mathcal C(E_n, L) &= \big [\! \cos^2 \delta_0 (E) \big ] \frac{\partial}{\partial E} \Big [ \widetilde F_{00}(E,L)^{-1}+\mathcal K^{(0)} (E) \Big ] \bigg \vert_{E = E_n(L)}.
\end{align}
We can see at this stage that the K-matrix takes the place of the scattering phase shift and $\widetilde F$ takes the place of the function $\phi$. This is made precise through the relations
\begin{align}
\label{eq:phidef}
\widetilde F_{00,00}(E,L)^{-1} \equiv \frac{16 \pi E}{p}\tan \phi(E,L) \,, \qquad \qquad \mathcal K^{(0)}(E) = \frac{16 \pi E}{p} \tan \delta_{0}(E) \,.
\end{align}
The conversion factor, $\mathcal{C}$, becomes
\begin{align}
\mathcal{C}^{\ell_{\sf{max}}=0}(E_n,L)=\big [\! \cos^2 \delta_0 (E) \big ] \frac{16 \pi E}{p} \frac{\partial}{\partial E} \big [\tan \phi(E, L) + \tan \delta_{0}(E) \big ] \bigg \vert_{E = E_n(L)} \,,
\end{align}
where we have used the fact that the energy levels satisfy $\mathbb M(E_n,L)=0$ to simplify. This can be shown to be equivalent to the the Lellouch-L{\"u}scher formalism given in eq.~\eqref{eq:LL_form}.

\section{Contribution from $\ell=4$ scattering}
\label{sec:l4}

As described in the introduction, due to the geometry of the finite volume, angular momentum is no longer a good quantum number. Instead, for vanishing total momentum, the system is defined by a particular irreducible representation (irrep) of the octahedral group. For $K \to \pi \pi$, the irrep of interest is the trivial irrep,\footnote{Pseudo-scalars, such as pions and kaons, transform trivially under rotations but incur a sign-flip under parity. However, only the two-pion state is of direct relevance to us and this is also even under parity. For this reason we consider the $A_1^+$ rather than the $A_1^-$ irrep.} $A_1^+$. As is well known, the $A_1^+$ also couples to higher angular-momentum states. After $\ell=0$, the next lowest angular-momentum state that contributes is $\ell = 4$ \cite{Luscher:1986pf,Luscher:1990ux,Luu:2011ep,Rummukainen:1995vs,Dudek:2012gj,Grabowska:2021xkp}. Naively, if we were to truncate at $\ell=4$, our angular momentum indices would run over $2\times 4+1=9$ additional values. However, as $A_1^+$ only couples to one specific $\ell=4$ state, we can perform a change of basis so that we end up with a $2\times2$ matrix space:
\begin{align}
\overline F_{\ell \ell'} = \langle A_1^+ , \ell \vert \widetilde F \vert A_1^+, \ell' \rangle \,,
\end{align}
where the $\vert A_1^+, \ell\rangle$ states are defined by
\begin{align}
\vert A_1^+, \ell = 0 \rangle&=\vert 0,0\rangle \,, \\
\vert A_1^+, \ell = 4 \rangle &= \frac{1}{2}\sqrt{\frac{5}{6}}\left(|4,4\rangle+|4,-4\rangle\right)+\frac{1}{2}\sqrt{\frac{7}{3}}|4,0\rangle \,,
\end{align}
with the right-hand side in the $\ell, m$ basis used for standard spherical harmonics.
In the basis that projects to $A_1^+$, the $\ell=4$-truncated $\mathbb M$ matrix becomes
\begin{align}
\label{eq:M_2x2}
\mathbb M(E, L) \equiv \begin{pmatrix} {\widetilde{\mathcal K} }^{(0)}(E) & 0 \\ 0 & {\widetilde{\mathcal K}}^{(4)}(E) \end{pmatrix}+
\begin{pmatrix}
\overline F_{00}(E,L) & \overline F_{04}(E,L) \\ \overline F_{40}(E,L) & \overline F_{44}(E,L)
\end{pmatrix}^{-1} \,.
\end{align}
In order to quantify the approximation of neglecting $\ell = 4$, we define $\Delta$ to be the relative difference between the $\ell=4$ and $\ell=0$ conversion factors:
\begin{align}
\Delta(E_n,L)\equiv \frac{ \mathcal C(E_n, L) - \mathcal C^{\ell_{\sf max} = 0}(E_n, L)}{\mathcal C^{\ell_{\sf max} = 0}(E_n, L)}.
\end{align}
Here $\mathcal{C}$ is assumed to be defined up to $\ell=4$, i.e.~with the form of $\mathbb M$ given in eq.~\eqref{eq:M_2x2}.

Expanding this to leading order in $\mathcal K^{(4)}$, also counting $\partial_E \mathcal K^{(0)}$ as suppressed relative to $\partial_E \overline F_{00}$, and once again using $\text{Det}[\mathbb M]=0$, we reach\footnote{In the remainder of this work we use $m$ rather than $m_\pi$ for the pion mass.}
\begin{multline}
\Delta(E_n, L)
= \bigg [ (2m)^9 \frac{\partial}{\partial E} \frac{\mathcal K^{(4)}(E) }{p^8} \bigg ]_{E=E_n} \Delta^{[\partial \mathcal K(4)]}(E_n, L)
+ \bigg [ (2m)^8 \frac{\mathcal K^{(4)}(E_n) }{p^8} \bigg ] \Delta^{[ \mathcal K(4)]}(E_n,L)
\\
+ \mathcal O \big [ (\mathcal K^{(4)})^2, \mathcal K^{(4)} \, \partial \mathcal K^{(0)} \big ] \,,
\label{eq:DeltaEexpanded}
\end{multline}
with
\begin{align}
\Delta^{[\mathcal K(4)]}(E,L) & \equiv \frac{1}{(2m)^8} \frac{1}{\partial_E \overline F_{00}(E, L)^{-1}}
\frac{ \overline F_{04}(E,L)^2}{\overline F_{00}(E, L)^2} \Bigg [ 2
\frac{ \partial_E \overline F_{04}(E,L)}{ \overline F_{04}(E,L)} - 2 \frac{\partial_E \overline F_{00}(E,L) }{ \overline F_{00}(E, L) } \\
& \hspace{40pt} + \frac{m^2}{E p^2} - \frac{2}{E p^2} \frac{1}{{\overline F}_{00}(E,L)} \frac{ m^2 {\overline F}_{00}(E,L) - E p^2 \partial_E {\overline F}_{00}(E,L)}{1 + \big [ 16 \pi E {\overline F}_{00}(E,L)/p \big ]^2} \bigg ] \,, \\[10pt]
\Delta^{[\partial \mathcal K(4)]}(E,L) & \equiv \frac{1}{(2m)^9} \frac{1}{\partial_E \overline F_{00}(E, L)^{-1}} \frac{ \overline F_{04}(E,L)^2}{\overline F_{00}(E,L)^2} \,.
\end{align}
Technically, $\Delta (E_n, L)$ should be evaluated at an energy $E_n$ that is a solution of the $\ell=4$ quantisation condition. However, this introduces corrections to the expansion of eq.~\eqref{eq:DeltaEexpanded} that are higher order in $\mathcal{K}^{(4)}$. Thus, when evaluating the expansion coefficient functions, $\Delta^{[\mathcal K(4)]} $ and $\Delta^{[\partial \mathcal K(4)]} $, it is sufficient to use the $\ell=0$ quantisation condition.

\begin{figure}[h]
\centering
\includegraphics[width=0.9\textwidth]{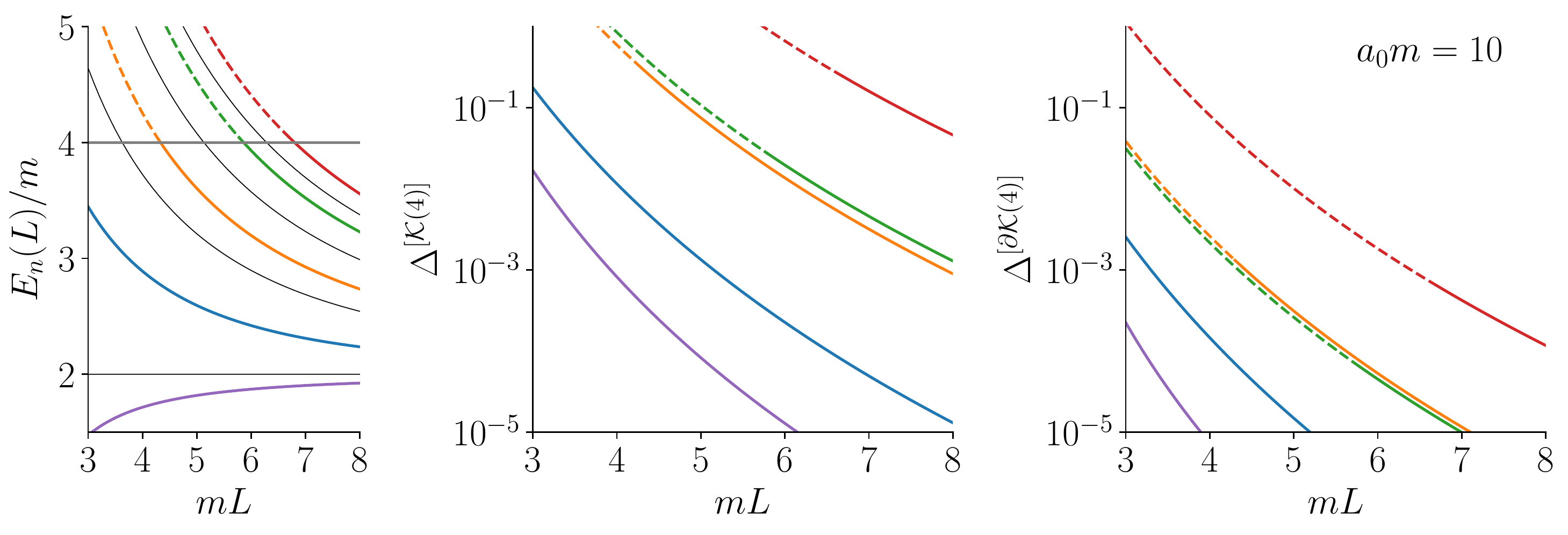}
\includegraphics[width=0.9\textwidth]{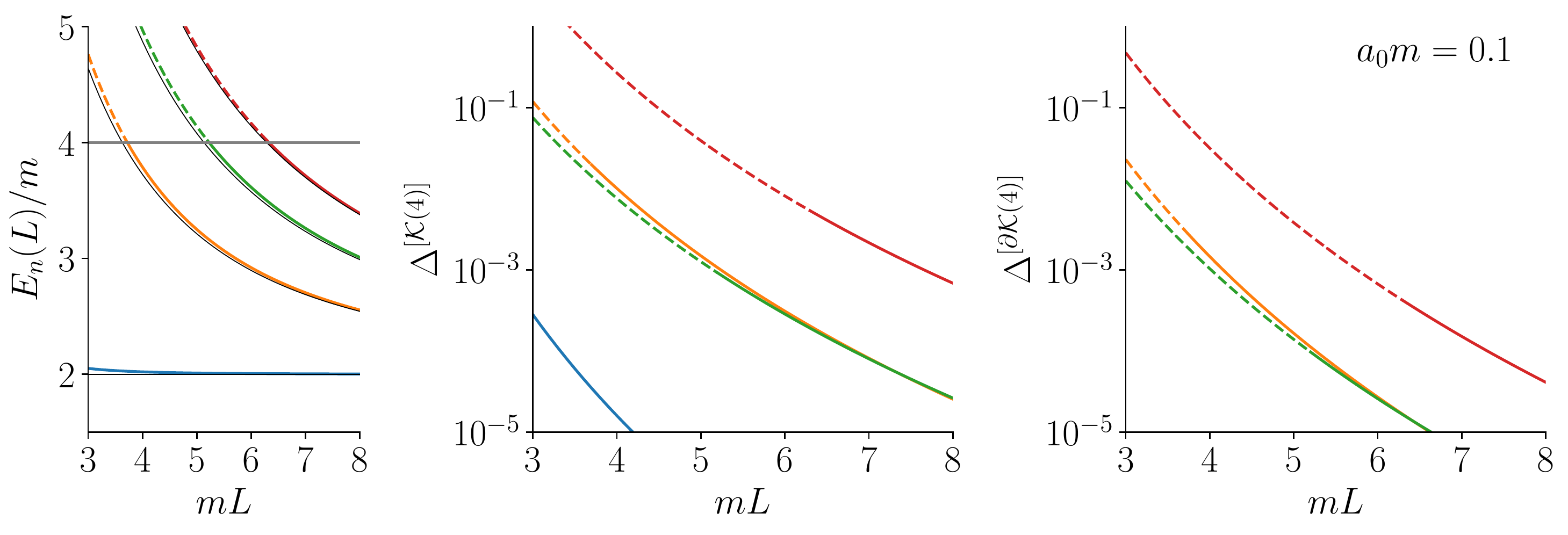}
\caption{$E_n(L)/m$, $\Delta^{[\mathcal K(4)]}(E_n(L), L)$ and $\Delta^{[\partial \mathcal K(4)]}(E_n(L), L)$ plotted verses $m L$ for various choices of the $S$-wave scattering length scattering length $m a_0$. The dashed segments indicate that $E_n(L) > 4 m$, implying that the formalism has neglected systematic uncertainties due to on-shell four-particle states. In the case of $ma_0=10$, a shallow boundstate arises, resulting in the lowest lying finite-volume energy shown in lilac.}
\label{fig:variousmal4th0}
\end{figure}

Although the functions $\Delta^{[\mathcal K(4)]}$ and $\Delta^{[\partial \mathcal K(4)]}$ do not have explicit dependence the dynamics of the theory, they are only meaningful at physical values of $E_n(L)$ and thus inherit a dependence on the interaction parameters through the interacting energy levels. In order to calculate these energy levels, we parameterise the $\ell = 0$ K-matrix with the scattering length $a_0$:
\begin{align}
\mathcal K^{(0)}(E) = - 16 \pi E a_0.
\end{align}
The results for the energy levels and $\Delta^{[\mathcal K(4)]}$ and $\Delta^{[\partial \mathcal K(4)]}$ evaluated at those energy levels, for various values of $a_0 m$, are shown in in figure~\ref{fig:variousmal4th0}. For some choices of $a_0 m$ we find the contamination from $\ell=4$ will be $\sim 10\%$ if the factorized $\ell = 4$ dynamics is order unity. Note that this is plausible, since the barrier factors are removed in the quantities representing the $\ell = 4$ interaction strength, so $  \frac{(2m)^8 \mathcal K^{(4)} }{p^8}$ is the quantity that would have to be order one. However, for other values of $a_0m$ we find that the effect is suppressed by many orders of magnitude. A detailed analysis for the parameters of physical pions and kaons will be presented in the full manuscript, to appear.

\section{The accidentally degenerate 9th energy level}\label{sec:accdeg}
The free energy levels have values
\begin{align}
(E^{\sf free}_{\vec n})^2/4=\left(\frac{2\pi}{L}\right)^2 \vec n^2+m^2 \,,
\end{align}
where $\vec n$ is a three-vector of integers. However, the energy level with $\vec n^2=9$ has 2 inequivalent solutions: $(3,0,0)$ and $(2,2,1)$, which cannot be transformed into each other using the octahedral group. The degeneracy is accidental as it is not enforced by the symmetries of the system.

\begin{figure}
\centering
\includegraphics[width=0.95\textwidth]{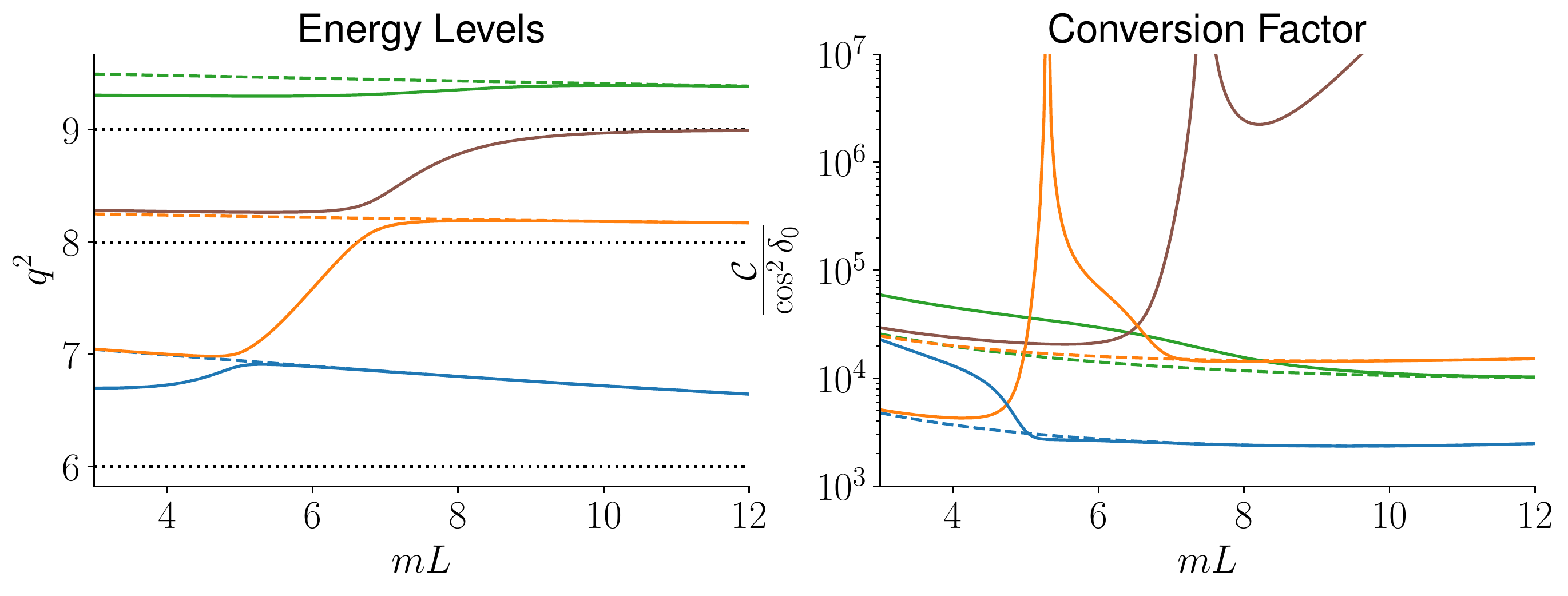}
\caption{Finite-volume energies (left panel) and conversion factors (right panel) for states in the vicinity of the accidentally degenerate state, with scattering parameters set to $m a_0=1.0$ and $ (m^9 a_4 ) =-0.0001$. In the case of the energies we plot $q^2 = L^2 (E^2/4 - m^2)/(2 \pi)^2$ on the vertical axis as this separates the curves more clearly. The black dotted lines are the free energy levels for reference, the coloured dashed lines are the solutions to the $\ell=0$ condition and the solid lines are the solutions to the $\ell=4$ condition.}
\label{fig:adenergycvals}
\end{figure}

In a weakly interacting theory, if one uses the $\ell=0$ quantisation condition, one finds a single interacting energy level slightly shifted from each of the free energy levels, including the 9th energy level. By contrast, truncating the quantisation condition at $\ell=4$, one finds that the accidentally degenerate level splits into two distinct energies. To asses the effect this has on extractions of the decay amplitudes, in figure~\ref{fig:adenergycvals} we plot the full conversion factor, as defined in eq.~\eqref{eq:conv_fact}. In particular, we do not expand in terms of $\ell=4$ quantities as we did in the previous section. In order to calculate the energy levels and conversion factors, we parameterise the 4th partial wave of the K-matrix as
\begin{align}
\mathcal{K}^{(4)}=-16\pi E p^8 a_4 \,,
\end{align}
and, for the 0th partial wave, we continue to use the same scattering length parameterisation as before. As shown in figure~\ref{fig:adenergycvals}, when the $\ell=4$ and $\ell=0$ energy levels agree, so do the respective conversion factors. However, when the $\ell=4$ energy levels deviate from the $\ell=0$ energy levels, the conversion factor varies rapidly and can even diverge.

The divergence can be explained as follows: A given finite-volume state can be viewed as a linear combination of infinite-volume states with different angular momenta \cite{Hansen:2012tf,Briceno:2014uqa} \footnote{This expression is only true in an inner product with an appropriate incoming state}:
\begin{align}
\sqrt{2 m_K} L^3 \, \langle E_n, L , A_1^+ \vert = \sum_{\ell}
\mathcal A_{\ell }(E_{n}, L)
\langle E_{n}, \pi \pi, \text{out}, \ell \vert \,,
\end{align}
where $\mathcal A_{\ell}(E_n, L)$ is proportional to the eigenvector of $\mathbb M$ with vanishing eigenvalue at the finite-volume energy solution. When this combination is applied to the state $\mathcal H_W |K\rangle$, only the $\ell=0$ infinite-volume state contributes and
\begin{equation}
\mathcal{C}(E_{n}, L) = \frac{L^6}{\vert \mathcal A_0(E_{n}, L) \vert^2} \,.
\end{equation}
The deviation of the $\ell=4$ energy levels from the $\ell=0$ energy levels corresponds to $\mathcal{A}_4$ dominating over $\mathcal{A}_0$. The divergence in $\mathcal{C}$ therefore occurs when the contribution from the $\ell=0$ infinite-volume state vanishes.

\section{Conclusion}

We have examined the standard approximation of truncating the $1\to2 $ Lellouch-L{\"u}scher relation, as it arises in calculations of $K \to \pi \pi$, by neglecting the effects of higher partial waves in $\pi \pi \to \pi \pi$ scattering, i.e.~approximating that the QCD scattering amplitude vanishes for $\ell \geq 4$. We have found that including the next lowest angular momentum state, $\ell=4$ in the rest frame, introduces a correction that can be as much as $10\%$, depending on the details of the finite-volume system and scattering parameters, but is often much smaller. A detailed analysis for physical pion scattering parameters will be given in the full manuscript, to appear.

Secondly, we have considered the phenomenological features that occur at the 9th energy level. The original Lellouch-L{\"u}scher formalism only applies below this energy level due to the two inequivalent non-interacting solutions with back-to-back momentum $p$ satisfying $L^2 p^2/(2 \pi)^2=9$. We find that including the $\ell=4$ contribution causes these energy levels to split. The Lellouch-L{\"u}scher conversion factor exhibits rapid variation in this region and can even diverge for specific choices of $L$, corresponding to a case with the finite-volume state is dominated by the $\ell = 4$ component.

\acknowledgments{M.T.H.~would like to thank Ra\'ul Brice\~no, Mattia Bruno, Fernando Romero-L\'opez and Steve Sharpe for useful discussions, and for previous and ongoing collaborations that helped to inspire this work. M.T.H.~is supported by UK Research and Innovation Future Leader Fellowship MR/T019956/1. M.T.H. and T.P.~are supported in part by UK STFC grant ST/P000630/1.}

\appendix
\section{Explicit form of $\widetilde{F}$}

The definition of $\widetilde F$ for vanishing total spatial momentum in the finite-volume frame is given by \cite{\DefOfF}
\begin{equation}
\label{eq:fdef}
\widetilde F_{\ell m, \ell' m'}(E, L) = \frac12 \lim_{\alpha \to 0^+} \bigg [ \frac{1}{L^3} \sum_{ \vec{k}} - \, \text{p.v.} \! \int \! \! \frac{d^3 \vec k}{(2 \pi)^3} \bigg ] \frac{\mathcal Y_{\ell m}(\vec k) \mathcal Y^*_{\ell' m'}(\vec k) \, e^{- \alpha (\vec k^2 - p^2)}}{(2 \omega_{ k})^2 (E - 2\omega_{ k} )} \,,
\end{equation}
where $p^2 = E^2/4-m^2$ and
\begin{align}
\mathcal Y_{\ell m}(\vec k) & = \sqrt{4 \pi} \vert \vec k \vert^{\ell} Y_{\ell m}(\hat { k}) \,, \\
\omega_{ k} & = \sqrt{\vec k^2 + m^2} \,.
\end{align}
Here the sum runs over three-vectors of the form
\begin{align}
\vec{k} = \frac{2\pi}{L}\vec{n}\,, \qquad \vec n \in \mathbb{Z}^3 \,,
\end{align}
and the p.v.~indicates that the principal-value pole prescription is used to make the integral well-defined in the vicinity of the pole. Finally, the exponential factor provides an ultraviolet regularisation of the sum and integral individually. These divergences cancel in the difference so that one can send $\alpha \to 0^+$ as indicated.

\bibliographystyle{JHEP}
\bibliography{refs.bib}

\providecommand{\href}[2]{#2}\begingroup\raggedright\begin{thebibliography}{10}

\bibitem{Lellouch:2000pv}
L.~Lellouch and M.~L{\"u}scher, \emph{{Weak transition matrix elements from
  finite volume correlation functions}},
  \href{https://doi.org/10.1007/s002200100410}{\emph{Commun. Math. Phys.}
  {\bfseries 219} (2001) 31}
  [\href{https://arxiv.org/abs/hep-lat/0003023}{{\ttfamily hep-lat/0003023}}].

\bibitem{Bai:2015nea}
{\scshape RBC, UKQCD} collaboration, \emph{{Standard Model Prediction for
  Direct CP Violation in $K \to \pi\pi$ Decay}},
  \href{https://doi.org/10.1103/PhysRevLett.115.212001}{\emph{Phys. Rev. Lett.}
  {\bfseries 115} (2015) 212001}
  [\href{https://arxiv.org/abs/1505.07863}{{\ttfamily 1505.07863}}].

\bibitem{Blum:2015ywa}
T.~Blum et~al., \emph{{$K \rightarrow \pi\pi$ $\Delta I=3/2$ decay amplitude in
  the continuum limit}},
  \href{https://doi.org/10.1103/PhysRevD.91.074502}{\emph{Phys. Rev. D}
  {\bfseries 91} (2015) 074502}
  [\href{https://arxiv.org/abs/1502.00263}{{\ttfamily 1502.00263}}].

\bibitem{Abbott:2020hxn}
{\scshape RBC, UKQCD} collaboration, \emph{{Direct CP violation and the $\Delta
  I=1/2$ rule in $K\to\pi\pi$ decay from the standard model}},
  \href{https://doi.org/10.1103/PhysRevD.102.054509}{\emph{Phys. Rev. D}
  {\bfseries 102} (2020) 054509}
  [\href{https://arxiv.org/abs/2004.09440}{{\ttfamily 2004.09440}}].

\bibitem{Lin:2001ek}
C.~Lin, G.~Martinelli, C.T.~Sachrajda and M.~Testa, \emph{{K --\ensuremath{>}
  pi pi decays in a finite volume}},
  \href{https://doi.org/10.1016/S0550-3213(01)00495-3}{\emph{Nucl. Phys. B}
  {\bfseries 619} (2001) 467}
  [\href{https://arxiv.org/abs/hep-lat/0104006}{{\ttfamily hep-lat/0104006}}].

\bibitem{Detmold:2004qn}
W.~Detmold and M.J.~Savage, \emph{{Electroweak matrix elements in the two
  nucleon sector from lattice QCD}},
  \href{https://doi.org/10.1016/j.nuclphysa.2004.07.007}{\emph{Nucl. Phys. A}
  {\bfseries 743} (2004) 170}
  [\href{https://arxiv.org/abs/hep-lat/0403005}{{\ttfamily hep-lat/0403005}}].

\bibitem{Kim:2005gf}
C.h.~Kim, C.T.~Sachrajda and S.R.~Sharpe, \emph{{Finite-volume effects for
  two-hadron states in moving frames}},
  \href{https://doi.org/10.1016/j.nuclphysb.2005.08.029}{\emph{Nucl. Phys.}
  {\bfseries B727} (2005) 218}
  [\href{https://arxiv.org/abs/hep-lat/0507006}{{\ttfamily hep-lat/0507006}}].

\bibitem{Christ:2005gi}
N.H.~Christ, C.~Kim and T.~Yamazaki, \emph{{Finite volume corrections to the
  two-particle decay of states with non-zero momentum}},
  \href{https://doi.org/10.1103/PhysRevD.72.114506}{\emph{Phys. Rev.}
  {\bfseries D72} (2005) 114506}
  [\href{https://arxiv.org/abs/hep-lat/0507009}{{\ttfamily hep-lat/0507009}}].

\bibitem{Meyer:2011um}
H.B.~Meyer, \emph{{Lattice QCD and the Timelike Pion Form Factor}},
  \href{https://doi.org/10.1103/PhysRevLett.107.072002}{\emph{Phys. Rev. Lett.}
  {\bfseries 107} (2011) 072002}
  [\href{https://arxiv.org/abs/1105.1892}{{\ttfamily 1105.1892}}].

\bibitem{Hansen:2012tf}
M.T.~Hansen and S.R.~Sharpe, \emph{{Multiple-channel generalization of
  Lellouch-L{\"u}scher formula}},
  \href{https://doi.org/10.1103/PhysRevD.86.016007}{\emph{Phys. Rev.}
  {\bfseries D86} (2012) 016007}
  [\href{https://arxiv.org/abs/1204.0826}{{\ttfamily 1204.0826}}].

\bibitem{Briceno:2012yi}
R.A.~Brice{\~n}o and Z.~Davoudi, \emph{{Moving multichannel systems in a finite
  volume with application to proton-proton fusion}},
  \href{https://doi.org/10.1103/PhysRevD.88.094507}{\emph{Phys. Rev.}
  {\bfseries D88} (2013) 094507}
  [\href{https://arxiv.org/abs/1204.1110}{{\ttfamily 1204.1110}}].

\bibitem{Bernard:2012bi}
V.~Bernard, D.~Hoja, U.G.~Mei{\ss}ner and A.~Rusetsky, \emph{{Matrix elements
  of unstable states}},
  \href{https://doi.org/10.1007/JHEP09(2012)023}{\emph{JHEP} {\bfseries 09}
  (2012) 023} [\href{https://arxiv.org/abs/1205.4642}{{\ttfamily 1205.4642}}].

\bibitem{Agadjanov:2014kha}
A.~Agadjanov, V.~Bernard, U.G.~Mei{\ss}ner and A.~Rusetsky, \emph{{A framework
  for the calculation of the $\Delta N \gamma^*$ transition form factors on the
  lattice}}, \href{https://doi.org/10.1016/j.nuclphysb.2014.07.023}{\emph{Nucl.
  Phys.} {\bfseries B886} (2014) 1199}
  [\href{https://arxiv.org/abs/1405.3476}{{\ttfamily 1405.3476}}].

\bibitem{Briceno:2014uqa}
R.A.~Brice{\~n}o, M.T.~Hansen and A.~Walker-Loud, \emph{{Multichannel 1
  $\rightarrow$ 2 transition amplitudes in a finite volume}},
  \href{https://doi.org/10.1103/PhysRevD.91.034501}{\emph{Phys. Rev.}
  {\bfseries D91} (2015) 034501}
  [\href{https://arxiv.org/abs/1406.5965}{{\ttfamily 1406.5965}}].

\bibitem{Feng:2014gba}
X.~Feng, S.~Aoki, S.~Hashimoto and T.~Kaneko, \emph{{Timelike pion form factor
  in lattice QCD}},
  \href{https://doi.org/10.1103/PhysRevD.91.054504}{\emph{Phys. Rev.}
  {\bfseries D91} (2015) 054504}
  [\href{https://arxiv.org/abs/1412.6319}{{\ttfamily 1412.6319}}].

\bibitem{Briceno:2015csa}
R.A.~Brice{\~n}o and M.T.~Hansen, \emph{{Multichannel 0 $\to$ 2 and 1 $\to$ 2
  transition amplitudes for arbitrary spin particles in a finite volume}},
  \href{https://doi.org/10.1103/PhysRevD.92.074509}{\emph{Phys. Rev.}
  {\bfseries D92} (2015) 074509}
  [\href{https://arxiv.org/abs/1502.04314}{{\ttfamily 1502.04314}}].

\bibitem{Muller:2020wjo}
F.~M\"uller and A.~Rusetsky, \emph{{On the three-particle analog of the
  Lellouch-L\"uscher formula}},
  \href{https://doi.org/10.1007/JHEP03(2021)152}{\emph{JHEP} {\bfseries 03}
  (2021) 152} [\href{https://arxiv.org/abs/2012.13957}{{\ttfamily
  2012.13957}}].

\bibitem{Hansen:2021ofl}
M.T.~Hansen, F.~Romero-L\'opez and S.R.~Sharpe, \emph{{Decay amplitudes to
  three hadrons from finite-volume matrix elements}},
  \href{https://doi.org/10.1007/JHEP04(2021)113}{\emph{JHEP} {\bfseries 04}
  (2021) 113} [\href{https://arxiv.org/abs/2101.10246}{{\ttfamily
  2101.10246}}].

\bibitem{Luscher:1990ux}
M.~Luscher, \emph{{Two particle states on a torus and their relation to the
  scattering matrix}},
  \href{https://doi.org/10.1016/0550-3213(91)90366-6}{\emph{Nucl. Phys. B}
  {\bfseries 354} (1991) 531}.

\bibitem{Rummukainen:1995vs}
K.~Rummukainen and S.A.~Gottlieb, \emph{{Resonance scattering phase shifts on a
  nonrest frame lattice}},
  \href{https://doi.org/10.1016/0550-3213(95)00313-H}{\emph{Nucl. Phys.}
  {\bfseries B450} (1995) 397}
  [\href{https://arxiv.org/abs/hep-lat/9503028}{{\ttfamily hep-lat/9503028}}].

\bibitem{Luscher:1986pf}
M.~L{\"u}scher, \emph{{Volume Dependence of the Energy Spectrum in Massive
  Quantum Field Theories. 2. Scattering States}},
  \href{https://doi.org/10.1007/BF01211097}{\emph{Commun. Math. Phys.}
  {\bfseries 105} (1986) 153}.

\bibitem{Luu:2011ep}
T.~Luu and M.J.~Savage, \emph{{Extracting Scattering Phase-Shifts in Higher
  Partial-Waves from Lattice QCD Calculations}},
  \href{https://doi.org/10.1103/PhysRevD.83.114508}{\emph{Phys. Rev. D}
  {\bfseries 83} (2011) 114508}
  [\href{https://arxiv.org/abs/1101.3347}{{\ttfamily 1101.3347}}].

\bibitem{Dudek:2012gj}
J.J.~Dudek, R.G.~Edwards and C.E.~Thomas, \emph{{S and D-wave phase shifts in
  isospin-2 pi pi scattering from lattice QCD}},
  \href{https://doi.org/10.1103/PhysRevD.86.034031}{\emph{Phys. Rev.}
  {\bfseries D86} (2012) 034031}
  [\href{https://arxiv.org/abs/1203.6041}{{\ttfamily 1203.6041}}].

\bibitem{Grabowska:2021xkp}
D.M.~Grabowska and M.T.~Hansen, \emph{{Analytic expansions of multi-hadron
  finite-volume energies: I. Two-particle states}},
  \href{https://arxiv.org/abs/2110.06878}{{\ttfamily 2110.06878}}.

\end{thebibliography}\endgroup

\end{document}